\documentclass[onecolumn,showpacs]{revtex4}

\usepackage{graphicx}
\usepackage{dcolumn}
\usepackage{bm}

\input epsf

\begin{document}

\title{\Large  A STUDY OF PHASE TRANSITION IN BLACK HOLE THERMODYNAMICS}

\author{\bf Ritabrata
Biswas\footnote{biswas.ritabrata@gmail.com}~And~Subenoy
Chakraborty.\footnote{schakraborty@math.jdvu.ac.in.com} }

\affiliation{Department of Mathematics, Jadavpur University, Kolkata-32, India}

\date{\today}

\begin{abstract}

This paper deals with five-dimensional black hole solutions in (a)
Einstein-Maxwell-Gauss-Bonnet theory with a cosmological constant
and (b)Einstein-Yang-Mills-Gauss-Bonnet theory for spherically
symmetric space time. In both the cases the possibility of phase
transition is examined and it is analyzed
  whether the phase transition is a Hawking-Page type phase transition or not.\\


\end{abstract}


\maketitle

\section{INTRODUCTION}
 The geometry of the event horizon of a black hole (BH) and its thermodynamics are
 nicely interrelated. BH temperature (known as Hawking temperature in literature)
 is proportional to surface gravity on the horizon while entropy is
  related to the area of the event horizon [by Hawking, S. W.(1975) and Bekenstein, J. D.(1973)] and they follow the
   first law of thermodynamics[ by Bekenstein, J. D.(1973)].
Although, the statistical nature of the BH thermodynamics is completely
unknown, yet the thermodynamical stability of the BH is characterized by the
sign of its heat capacity. In fact, a BH is said to be thermodynamically unstable
[as Schwarzschild BH (SBH)] if heat capacity$ < 0$ while if heat capacity changes
sign in the parameter space such that it diverges [by Hut, P.(1977)] then ordinary thermodynamics
 tells us that it is a second order phase transition [by Davies, P. C. W.(1977,1977 and 1989)]. Further for extremal BH
 there exists a critical point and phase transition takes place from an extremal BH to
 its non-extremal counter part.

Gross et. al. [by Gros, D.J. et al (1982)] had shown that the SBH having negative specific heat is in an unstable
equilibrium with $T$, the temperature of the heat reservoir. In fact due to small fluctuations
 the SBH will either decay (to hot flat space) or grow absorbing thermal radiation in the heat
 reservoir [by York, J. W.(1986)]. However it is possible to have a thermodynamically stable BH having positive specific
  heat through a phase transition from thermal radiation [by Myung, Y. S.(2008 and 2007)].In this context, a mechanism of phase
   transition was introduced by Hawking and Page [Hawking, S. W. et al(1983)] showing the transition between
   thermal AdS space and Schwarzschild-AdS (SAdS) BH [Myung, Y. S.(2008 and 2008), Brown, J. D. et al (1994) and Witten, E.(1983)].

In this work, we will analyze the thermodynamical quantities
namely the free energy, specific heat etc. for possibility of a
phase transition for both Einstein-Maxwell-Gauss-Bonnet (EMGB)- BH
and Einstein-Yang-Mills-Gauss-Bonnet (EYMGB) BH.

\section{Phase Transition of EMGB BHs}

The action in five dimensional space time $(M,~g_{\mu\nu})$
 that represents Einstein-Maxwell theory with a Gauss-Bonnet term and a
cosmological constant has the expression [Boulware, D. G. et al.(1985), Wilthire, D. L.(1986 and 1988), Thibeault, M. et al(2006)]($8\pi G=1=c$)
\begin{equation}
S=\frac{1}{2}\int_{M}
d^{5}x\sqrt{-g}[R-2\Lambda-\frac{1}{4}F_{\mu\nu}F^{\mu\nu} +\alpha
R_{GB}]
\end{equation}

where
$R_{GB}=R^{2}-4R_{\alpha\beta}R^{\alpha\beta}+R_{\alpha\beta\gamma\delta}R^{\alpha\beta\gamma\delta}$,
 is the Gauss-Bonnet term, $\alpha$ is the GB coupling parameter
having dimension $(length)^{2}$ ( $\alpha^{-1}$ is related to
string tension in heterotic super string theory), $\Lambda$ is the
cosmological constant and
$F_{\mu\nu}=\left(\partial_{\mu}A_{\nu}-\partial_{\nu}A_{\mu}\right)$
is the usual electromagnetic field tensor with $A_{\mu}$, the
vector potential.Now variation of this action with respect to the
metric tensor and $F_{\mu\nu}$gives the modified Einstein field
equations and Maxwell's equations
\begin{equation}
G_{\mu\nu}-\alpha H_{\mu\nu}+\Lambda g_{\mu\nu}=T_{\mu\nu}
\end{equation}

and
\begin{equation}
\nabla_{\mu}F_{\nu}^{\mu}=0
\end{equation}

where $H_{\mu\nu}$ is the Lovelock tensor given by
\begin{equation}
H_{\mu \nu}=\frac{1}{2}g_{\mu \nu}(R_{\alpha \beta \gamma
\delta}R^{\alpha \beta \gamma \delta}-4R_{\alpha \beta} R^{\alpha
\beta}+R^{2})-2RR_{\mu \nu}+4R^{\lambda}_{\mu}R_{\lambda
\nu}+4R^{\rho \sigma}R_{\mu \rho \nu \sigma}-R^{\alpha \beta
\gamma}_{\mu}R_{\nu \alpha \beta \gamma}
\end{equation}

and
\begin{equation}
T_{\mu\nu}=2F^{\lambda}_{\mu}F_{\lambda\nu}-\frac{1}{2}F_{\lambda\sigma}F^{\lambda\sigma}g_{\mu\nu}
\end{equation}

is the electromagnetic stress tensor.

(Note that  the modified Einstein field equations (2) do not
contain any derivatives of the curvature terms and hence the field
equations remain second order).

 If the manifold $M$ is chosen to be five dimensional
spherically symmetric space-time having the line element
\begin{equation}
ds^{2}=-B(r)dt^{2}+B^{-1}(r)dr^{2}+r^{2}\left(d\theta_{1}^{2}+sin^{2}\theta_{1}(d\theta_{2}^{2}+\sin
^{2}\theta_{2}d\theta_{3}^{2})\right)
\end{equation}
with

$0 \leq \theta_{1}~~,~~\theta_{2}\leq \pi~~,~~0\leq\theta_{3}\leq
2\pi$,

 then solving the above field equations one obtains [Hawking, S. W. (1975) and Wilthire, D. L.(1986 and 1988)]
\begin{equation}
B(r)=1+\frac{r^{2}}{4\alpha}-\frac{r^{2}}{4\alpha}\sqrt{1+\frac{16M\alpha}{\pi
r^{4}}-\frac{8Q^{2}\alpha}{3r^{6}}+\frac{4\Lambda\alpha}{3}}
\end{equation}

Here in an orthonormal frame the non-null components of the
electromagnetic tensors are $ F_{\hat{t}\hat{r}}=
-F_{\hat{r}\hat{t}} =\frac{Q}{4 \pi r^{3}}$. Note that in the
limit $\alpha \rightarrow 0$ one may recover the Einstein-Maxwell
solution with a cosmological constant. Further, in the limit with
$\Lambda=0$ we have the five-dimensional
Reissner-Nordstr$\ddot{o}$m Solution and hence the parameters
$M(>0)$ and $Q$ can be identified as the mass and charge
respectively. Moreover, for the solution $(7)$ to be well defined,
the radial coordinate $r$ must have a minimum value ($r_{min}$) so
that the expression within the square root is positive
definite,i.e, the solution $(7)$ is well defined for $r>r_{min}$
where $r_{min}$ satisfies

$$1+\frac{16m\alpha}{\pi
r_{min}^{4}}-\frac{8Q^{2}\alpha}{3r_{min}^{6}}+\frac{4\Lambda\alpha}{3}=0$$

The surface $r=r_{min}$ corresponds to a curvature singularity.
However,depending on the values of the parameters this singular
surface may be surrounded by the event horizon (having radius
$r_{h}$ such that $B(r_{h})=0$)and the solution $(7)$ describes a
black hole solution known as EMGB BH. On the other hand, if no
event horizon exists then the above solution represents a naked
singularity.

\subsection{Calculation of thermodynamic quantities }

We can write the mass parameter for EMGB BH as(from $B(r_{h})=0$)
\begin{equation}
M=\pi \left[\frac{Q^{2}}{6}r_{h}^{-2}+\alpha+\frac{1}{2}r_{h}^{2}-\frac{\Lambda}{12}r_{h}^{4}\right]
\end{equation}
The surface area of the event horizon is given by, $A=2 \pi^{2}r_{h}^{3}$ and hence the entropy of the black hole by [Boulware, D. G. (1985)] takes the form
$$S=\frac{K_{B}A}{4 G \hbar}=\frac{K_{B}\pi^{2}}{2 G \hbar}r_{h}^{3}$$
Now choosing $\hbar=1$ and the Boltzmann constant appropriately, we have,
\begin{equation}
S=r_{h}^{3}
\end{equation}
To find the thermodynamic quantities we follow Chakraborty et.al.(2008) and Biswas et. al.(2009)
and find
\begin{equation}
\bullet~~ Hawking~Temperature~ (T_{H})~=-\frac{\pi}{9r_{h}^{5}}\left(Q^{2}-3r_{h}^{4}+\Lambda r_{h}^{6}\right)
\end{equation}

\begin{equation}
\bullet~~ Specific~Heat (c_{Q})~=T\left(\frac{\partial S}{\partial T}\right)_{Q}=3r_{h}^{3}\frac{\left(Q^{2}-3r_{h}^{4}+\Lambda r_{h}^{6}\right)}{\left(-5Q^{2}+3r_{h}^{4}+\Lambda r_{h}^{6}\right)}
\end{equation}

\begin{equation}
\bullet~~ Free~Energy~ (F)~=M-T_{H}.S=\frac{\pi}{36}\left(10 Q^{2}r_{h}^{-2}+36\alpha+6 r_{h}^{2}+\Lambda r_{h}^{4}\right)
\end{equation}

\subsection{Graphical analyisis of phase transition}
The figures $1-3$ shows the variation of the above thermodynamical
quantities with the variation of the radius of the event horizon
while figure 4 represents the functional dependenceof $F$ over
$T_{H}$.

In each figure there are three set of diagrams corresponding to
$Q=0.5,~~1~~ and~~2$ respectively. In figure $1(a)$ we have
plotted $T_{H}$ for $\Lambda=-1$ with $Q=0.5,~~1~~ and~~2$
represented by solid line, dashed line and dotted line
respectively. Similar are the figures $1(b)$ and 1(b) and 1(c) for
$\lambda=0~~and~~\lambda=+1$ respectively.

In the figures $1(a)-(c)$ $T_{H}$ has a maximum (and also a
minimum for $\lambda=-1$) at some finite $r_{h}$ while
asymptotically, $T_{H}\rightarrow +\infty$ for $\lambda=-1,~~
T_{H}\rightarrow 0$ for $\lambda=0$ and $T_{H}\rightarrow -\infty$
for $\lambda=+1$ respectively.

The graphs of $c_{Q}$ in figures $2(a)-2(c)$ show some distinct
features. In figure $(2a)$ for $\lambda=-1$ and $Q=0.5$ (the solid
line) $c_{Q}$ changes sign at two finite value of $r_{h}$, i.e.,
$c_{Q}$ changes form positive value to negative value and then
again $c_{Q}$ becomes positive. So the black hole is a stable one
for small $r_{h}$ and then it becomes unstable at intermediate
$r_{h}$ and finally it again becomes stable. So it is expected
that there are two phase transitions for the evolution of the
black hole. But for $Q=1$ and $Q=2$ (described by dash and dot
lines respectively) $c_{Q}$ always remains positive and hence
there is no question of any phase transition. For $\Lambda=0$ in
figure $2(b)$ all three graphs have the similar behavior -- the BH
changes from stable phase to an unstable one by changing sign of
$c_{Q}$ at some finite $r_{h}$. In figure $2(c)$ for $\Lambda=+1$,
the graphs of $c_{Q}$ for $Q=0.5$ and $Q=1$ cross the $r_{h}$-axis
twice and there is a transition from stable BH to an unstable one
and then again the BH will be stable. But for $Q=2$ the BH was
initially unstable and finally becomes a stable one through a
phase transition.

The variation of $F$ over $r_{h}$ has been presented in figures
$3(a)-(c)$. In all the three sets of figures the behavior of $F$
does not change significantly for different values of $Q$. For
$\Lambda=-1$ (in fig. $3(a)$) initially $F$ is positive for small
$r_{h}$ and then decreases to negative values and falls sharply to
$-\infty$. But for other two values of $\Lambda$ (i.e, $\Lambda=0$
and $\Lambda=+1$) $F$ remains positive through with a minima at
finite $r_{h}$.

The graphical presentation of the free energy $F$ over the Hawking
temperature $T_{H}$ are shown in figures $4(a)-(c)$ for $Q=0.5$.
In all the figures there are double points (cusp-type) (three in
fig $4(a)$) which possibly indicate a Hawking-Page type phase
transition.
\begin{figure}
\includegraphics[height=1.66in, width=2in]{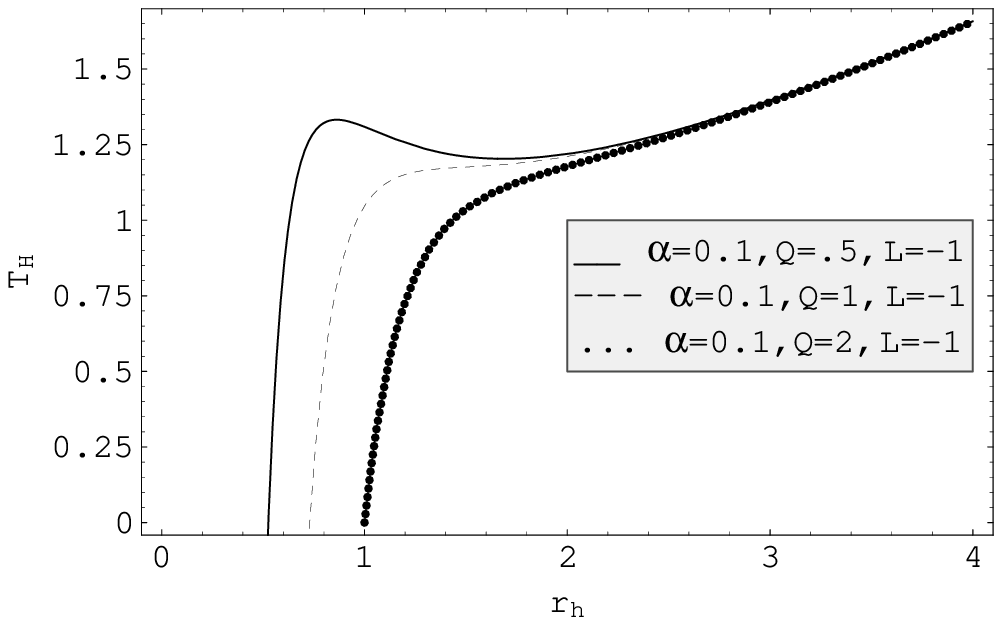}
~~\includegraphics[height=1.66in, width=2in]{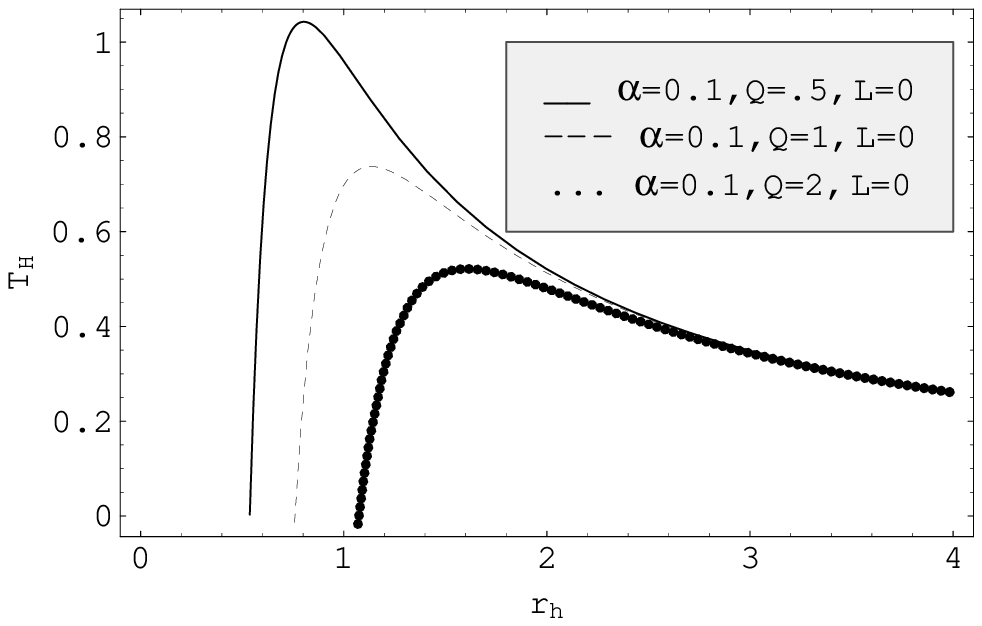}
~~\includegraphics[height=1.66in, width=2in]{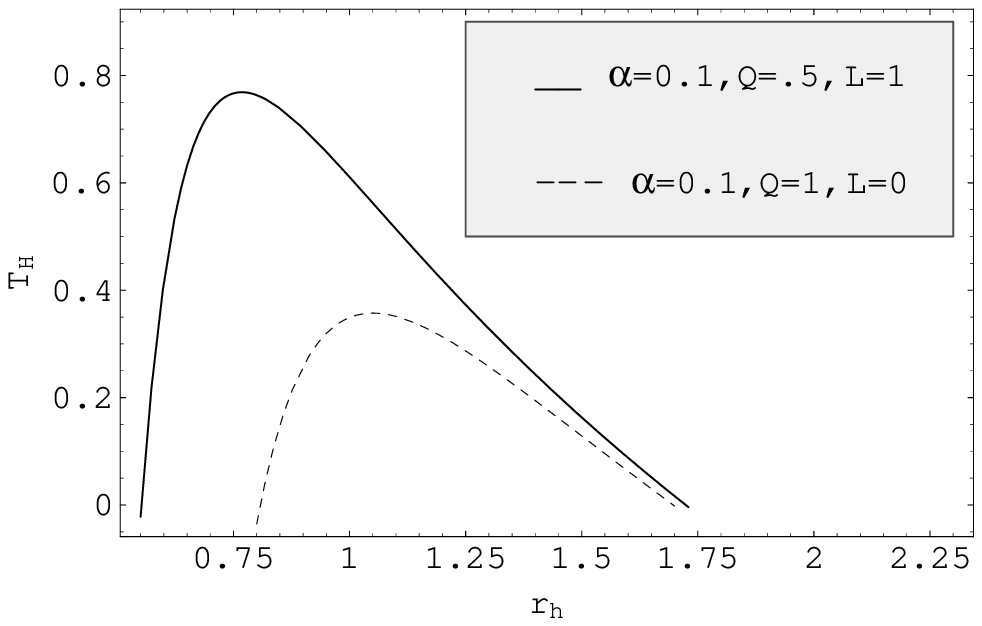}\\

Fig.1(a)~~~~~~~~~~~~~~~~~~~~~~~~~~~~~~~~~~~~~~Fig.1(b)~~~~~~~~~~~~~~~~~~~~~~~~~~~~~~~~~~~~~~~~Fig.1(c)\\

\vspace{5mm} Fig. 1(a), 1(b) and 1(c) show the variation of
$T_{H}$ with $r_{h}$ for $\Lambda(mentioned~~ as~~ L~~ in~~ the~~
figure)=-1,~0~and~1$ respectively for $\alpha =0.1$ in the case of
EMGB black hole. \hspace{1cm}
 \vspace{6mm}
\end{figure}

\begin{figure}
\includegraphics[height=1.66in, width=2in]{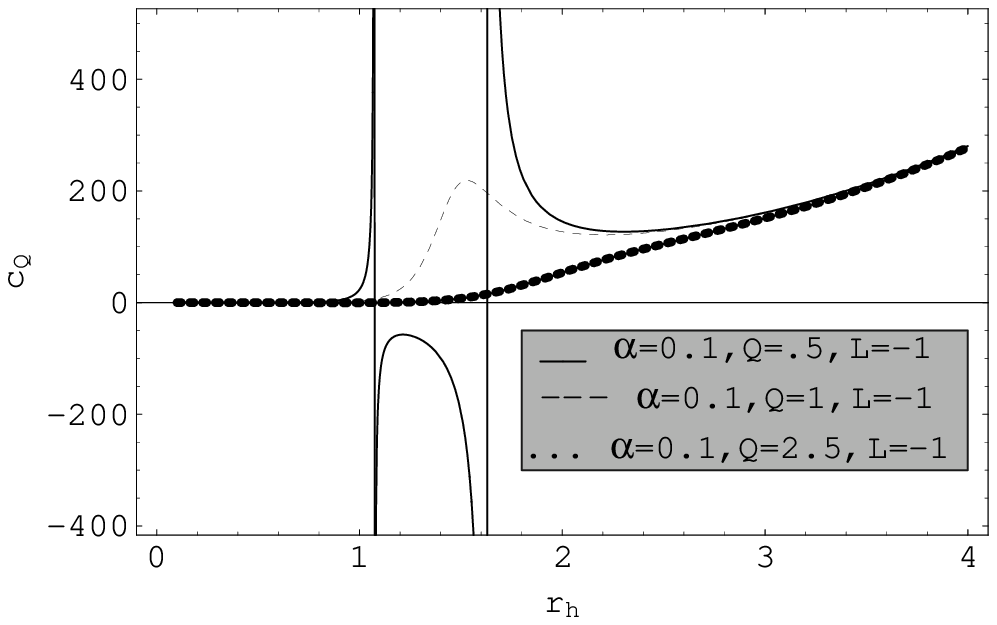}
~~\includegraphics[height=1.66in, width=2in]{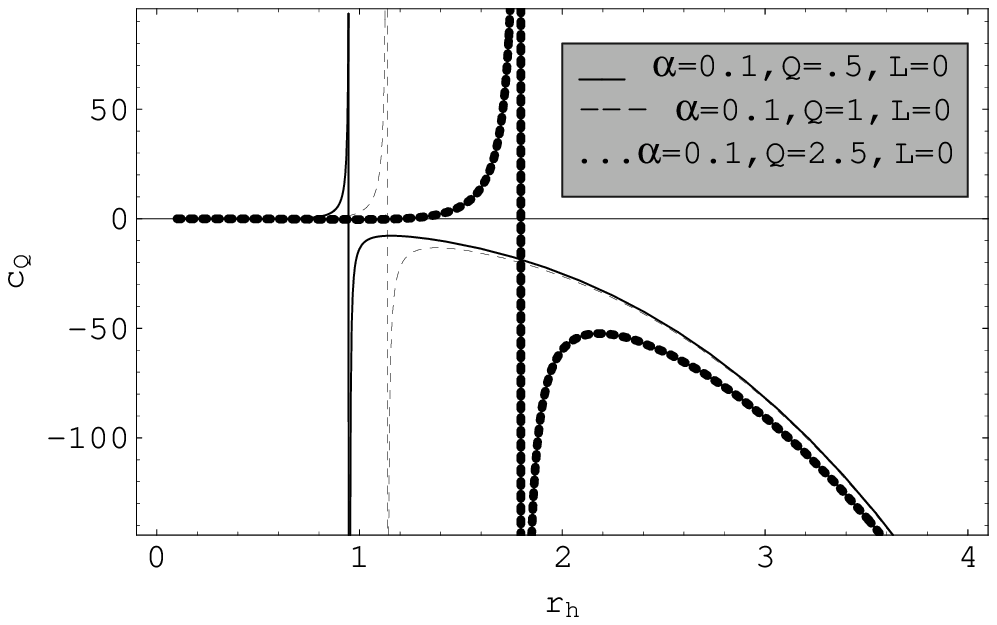}
~~\includegraphics[height=1.66in, width=2in]{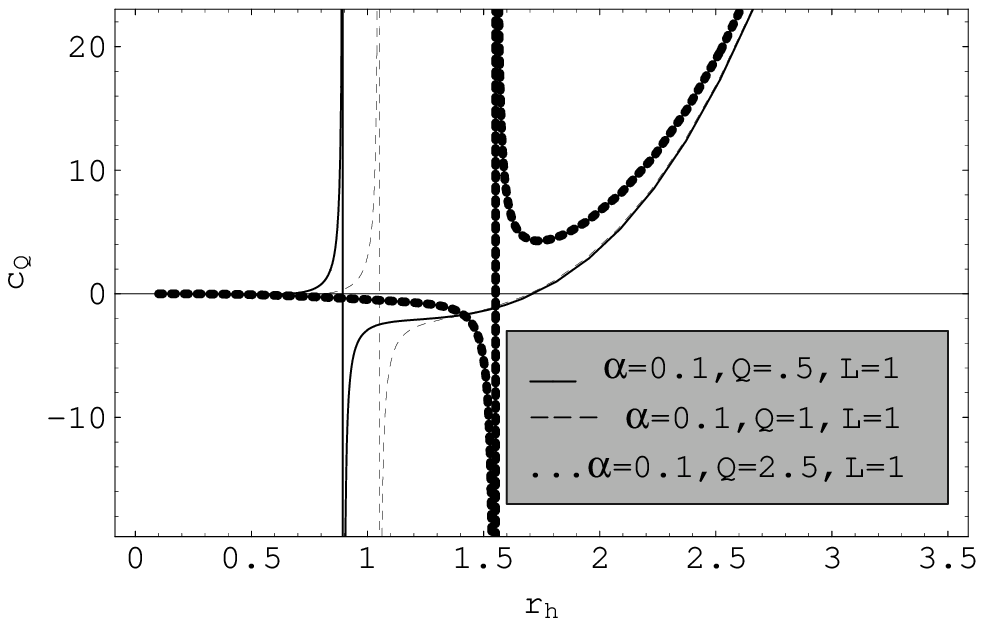}\\

Fig.2(a)~~~~~~~~~~~~~~~~~~~~~~~~~~~~~~~~~~~~~~Fig.2(b)~~~~~~~~~~~~~~~~~~~~~~~~~~~~~~~~~~~~~~~~Fig.2(c)\\

\vspace{5mm} Fig. 2(a), 2(b) and 2(c) show the variation of $c_{Q}$ with $r_{h}$ for $\Lambda=-1,~0~and~1$ respectively for $\alpha =0.1$
in the case of EMGB black hole.
\hspace{1cm}
 \vspace{6mm}
\end{figure}

\begin{figure}
\includegraphics[height=1.66in, width=2in]{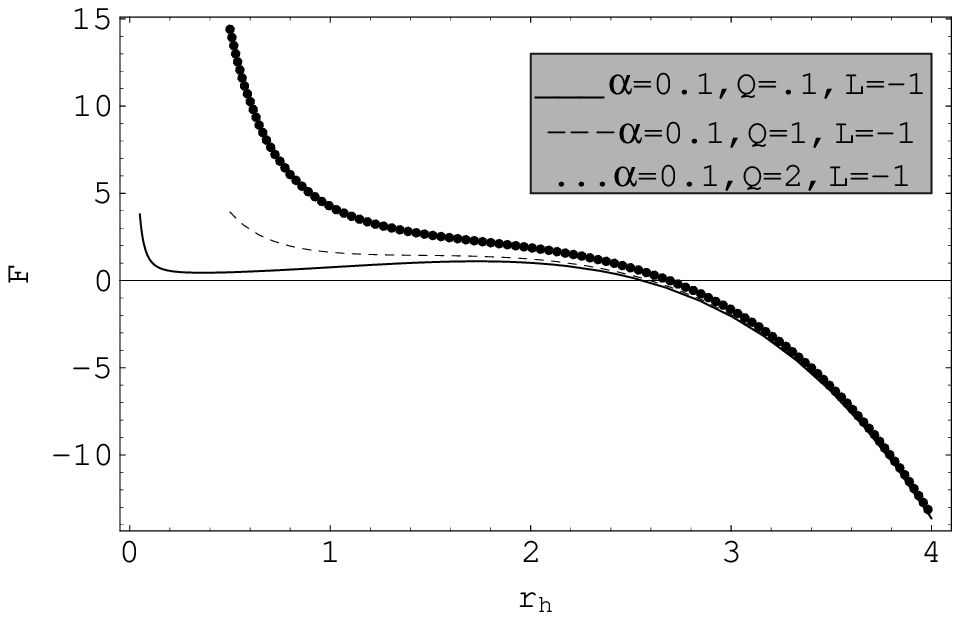}
~~\includegraphics[height=1.66in, width=2in]{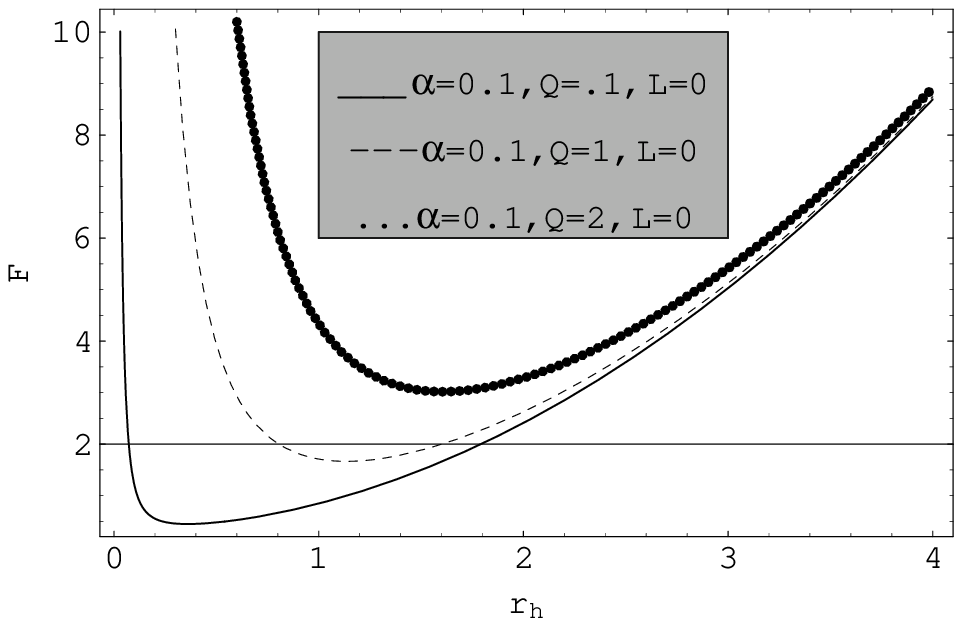}
~~\includegraphics[height=1.66in, width=2in]{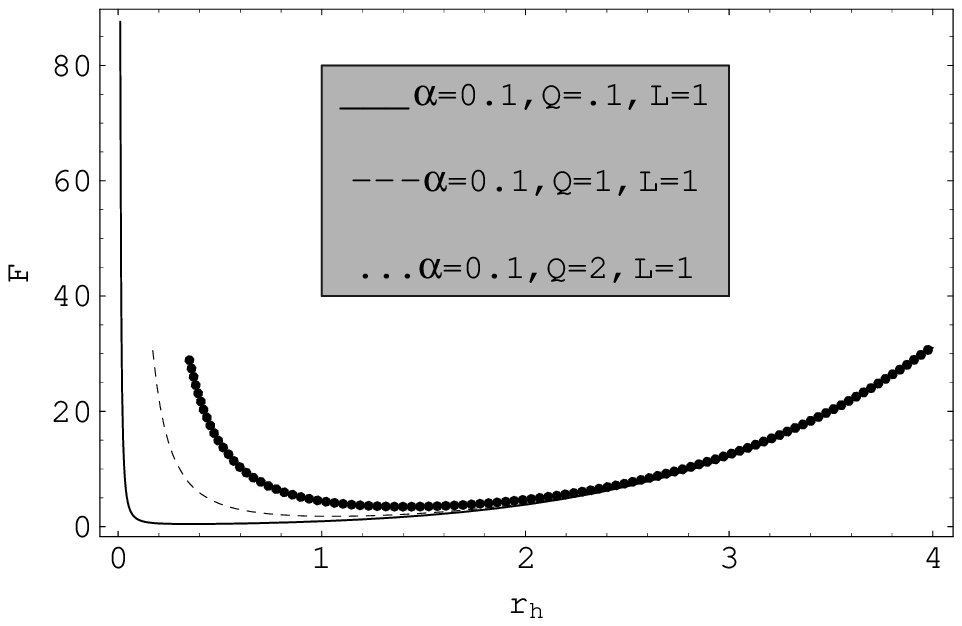}\\

Fig.3(a)~~~~~~~~~~~~~~~~~~~~~~~~~~~~~~~~~~~~~~Fig.3(b)~~~~~~~~~~~~~~~~~~~~~~~~~~~~~~~~~~~~~~~~Fig.3(c)\\

\vspace{5mm} Fig. 3(a), 3(b) and 3(c) show the variation of $F$ with $r_{h}$ for $\Lambda=-1,~0~and~1$ respectively for $\alpha =0.1$
in the case of EMGB black hole.
\hspace{1cm}
 \vspace{6mm}
\end{figure}

\begin{figure}
\includegraphics[height=1.66in, width=2in]{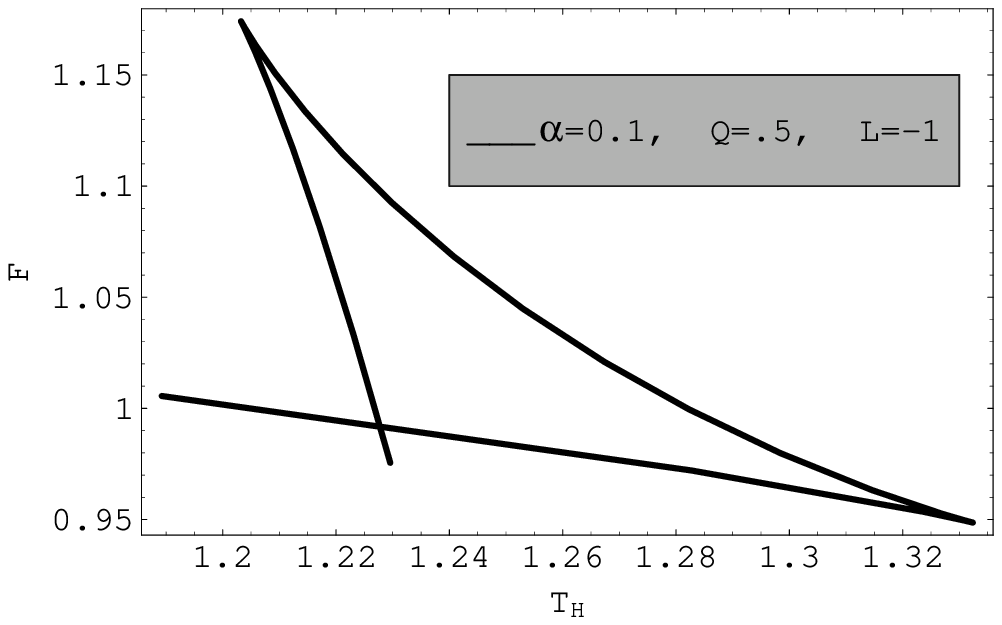}
~~\includegraphics[height=1.66in, width=2in]{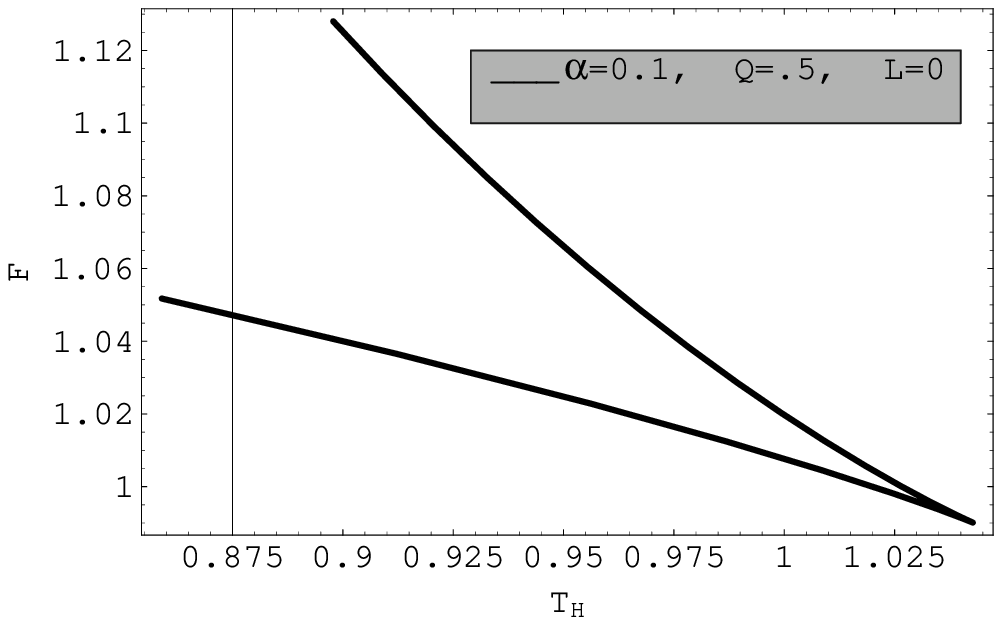}
~~\includegraphics[height=1.66in, width=2in]{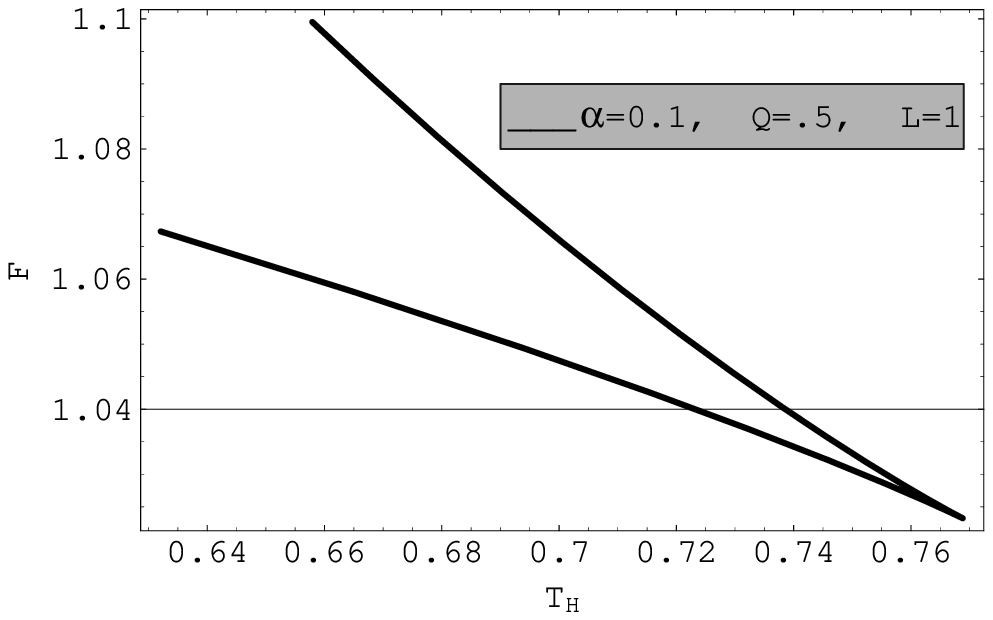}\\

Fig.4(a)~~~~~~~~~~~~~~~~~~~~~~~~~~~~~~~~~~~~~~Fig.4(b)~~~~~~~~~~~~~~~~~~~~~~~~~~~~~~~~~~~~~~~~Fig.4(c)\\

\vspace{5mm} Fig. 4(a), 4(b) and 4(c) show the variation of $F$ with $T_{H}$ for $\Lambda=-1,~0~and~1$ respectively for $\alpha =0.1$
in the case of EMGB black hole.
\hspace{1cm}
 \vspace{6mm}
\end{figure}

\section{Phase transition of EYMGB BHs}

Mazharimousavi and Halilsoy (2007) have recently obtained a 5-D
spherically symmetric solution in EYMGB theory. The metric ansatz
for 5-D spherically symmetric space-time is chosen as
\begin{equation}
ds^{2} = -U(r)dt^{2}+\frac{dr^{2}}{U(r)}+r^{2}d\Omega_{3}^{2}
\end{equation}

where they have expresed the metric on unit three sphere
$d\Omega_{3}^{2}$ in terms of Euler angles as [Mazharimousavi and Halilsoy (2007)]
\begin{equation}
d\Omega_{3}^{2}
=\frac{1}{4}(d\theta^{2}+d\phi^{2}+d\psi^{2}-2\cos\theta d\phi
d\psi )
\end{equation}

with ~~~$\theta ~\epsilon~[0,\pi],~~~ (\phi,~\psi)~\epsilon~[0,
2\pi]$.

For the Yang-Mills field the energy momentum tensor is given by,
\begin{equation}
T_{\mu\nu}=2F_{\mu}^{i \alpha}F^{i}_{\nu \alpha}-\frac{1}{2}g_{\mu
\nu}F^{i}_{\alpha \beta}F^{i \alpha \beta}
\end{equation}

where $F^{i}_{\alpha \beta}$ is the Yang-Mills field 2-forms such
that $F^{i}_{\alpha \beta}F^{i \alpha
\beta}=\frac{6Q^{2}}{r^{4}}$, $Q$ the only non-zero gauge charge.
The modified Einstein equations in EYMGB theory are
\begin{equation}
G_{\mu \nu}-\alpha H_{\mu \nu}=T_{\mu\nu}
\end{equation}

Where $G_{\mu \nu}$ is the usual Einstein tensor in 5-D,
$T_{\mu\nu}$ is the energy-momentum tensor given by equation
$(6)$, $\alpha$, the GB coupling parameter is chosen to be
positive in the heterotic string theory and the Lovelock tensor
has the expression given by equation $(4)$.

Now solving the non vanishing components of the field equations we
have [Mazharimousavi and Halilsoy (2007)]
\begin{equation}
U(r)=1+\frac{r^{2}}{4\alpha}\pm\sqrt{\left(\frac{r^{2}}{4\alpha}\right)^{2}+\left(1+\frac{m}{2
\alpha}\right)+\frac{Q^{2} \ln r }{\alpha}}
\end{equation}

with $m$ as the constant of integration. Now as $\alpha
\rightarrow 0$

\begin{equation}
U(r) \rightarrow 1-\frac{m}{r^{2}}-\frac{2 Q^{2} \ln r}{r^{2}}
\end{equation}

provided negative branch is considered. The metric coefficient
$U(r)$ in $(18)$ is identical to that of the Einstein-Yang-Mills
solution and hence $'m'$ is interpreted as the mass of the system.
In the equation $(17)$ for $U$ the expression within the square
root is positive definite for $\alpha>0$ while the geometry has a
curvature singularity at the surface $r=r_{min}$ for $\alpha<0$ .
Here $r_{min}$ is the minimum value of the radial coordinate such
that the function under the square root is positive. Moreover,
depending on the values of the parameters $(m,Q,\alpha)$, the
singular surface can be surrounded by an event horizon with radius
$r_{h}$ so that the space-time given by equation $(1)$ represents
a black hole. However if no event horizon exists, then there will
be naked singularity.

    Now the metric described by equation $(1)$ and $(17)$ has a singularity
at the greatest real and positive solution ($r_{s}$)of the
equation
\begin{equation}
\frac{r^{4}}{16 \alpha^{2}}+\left(1+\frac{m}{2
\alpha}\right)+\frac{Q^{2} \ln r}{\alpha}=0
\end{equation}

Note that if equation $(19)$ has no real positive solution then
the metric diverges at $r=0$. However, the singularity is
surrounded by the event horizon $r_{h}$, which is the positive
root of (the larger one if there are two positive real roots)

\begin{equation}
r^{2}-m-2 Q^{2} \ln r=0
\end{equation}

Thus if $r_{s}<r_{h}$ then the singularity will be covered by the
event horizon. While the singularity will be naked for $r_{s}\geq
r_{h}$. In this connection one may note that the event horizon is
independent of the coupling parameter $\alpha$.

\subsection{Calculation of the thermodynamical quantities}

We will discuss the thermodynamics [Aman, J. et. al.(2003)] of the black hole
described above.
 As the event horizon $r_{h}$ satisfies $(20)$ so we have,
\begin{equation}
\bullet~~~ m=r_{h}^{2}-2 Q^{2} \ln r_{h}
\end{equation}
As before , with proper choice of units $S=r_{h}^{3}$ and we find,
\begin{equation}
\bullet~~~ T_{H}=\left(\frac{\partial m}{\partial S}\right)_{Q}=\frac{2}{3}r_{h}^{-3}\left[r_{h}^{2}-Q^{2}\right]
\end{equation}
\begin{equation}
\bullet~~~ c_{Q}=\frac{3 r_{h}^{3}\left(r_{h}^{2}-Q^{2}\right)}{\left(3 Q^{2}-r_{h}^{2}\right)}
\end{equation}
\begin{equation}
\bullet~~~ F=\frac{1}{3}r_{h}^{2}+\frac{2}{3}Q^{2}\left[1-3\ln r_{h}\right]
\end{equation}

\subsection{Graphical interpretation of phase transition}

\begin{figure}
\includegraphics[height=2in, width=2in]{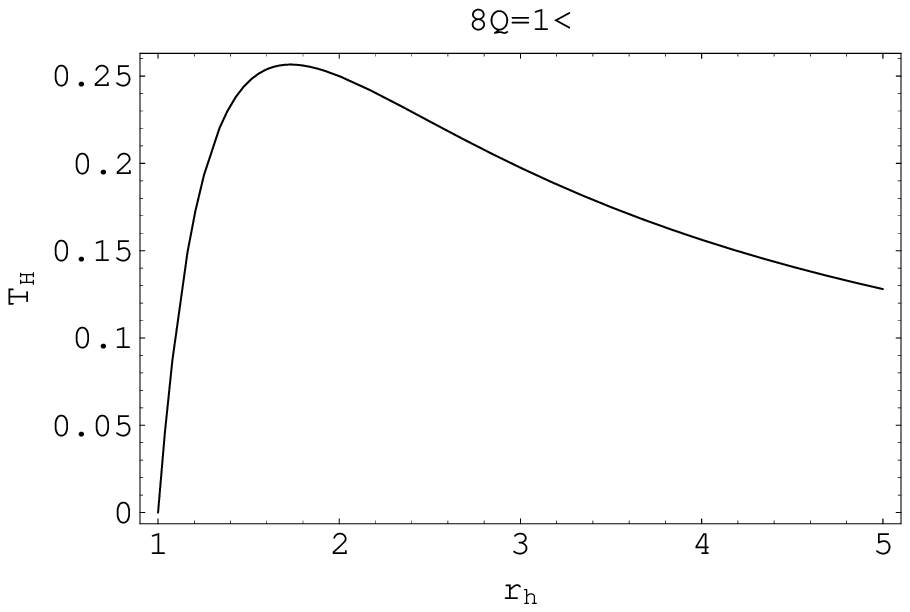}
~~\includegraphics[height=2in, width=2in]{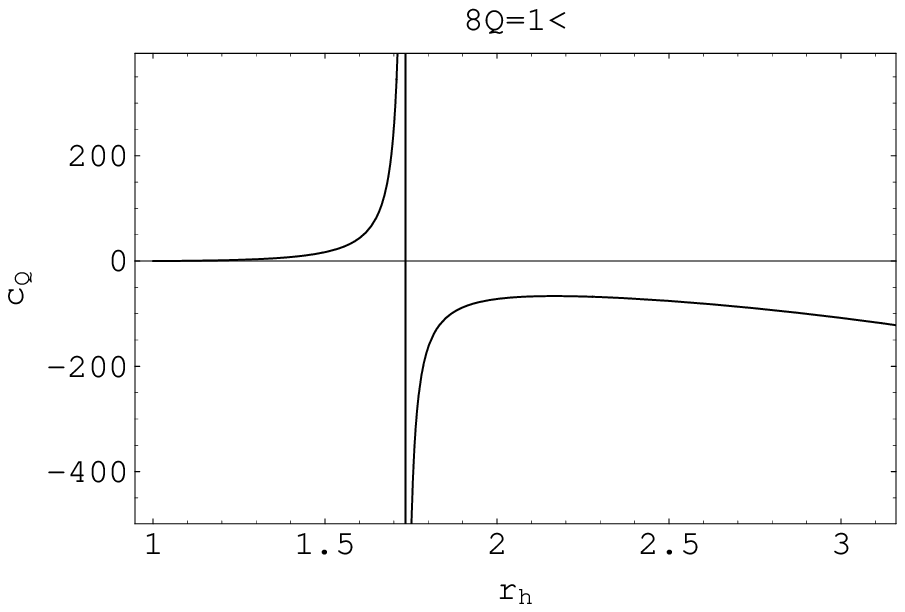}
~~\includegraphics[height=2in, width=2in]{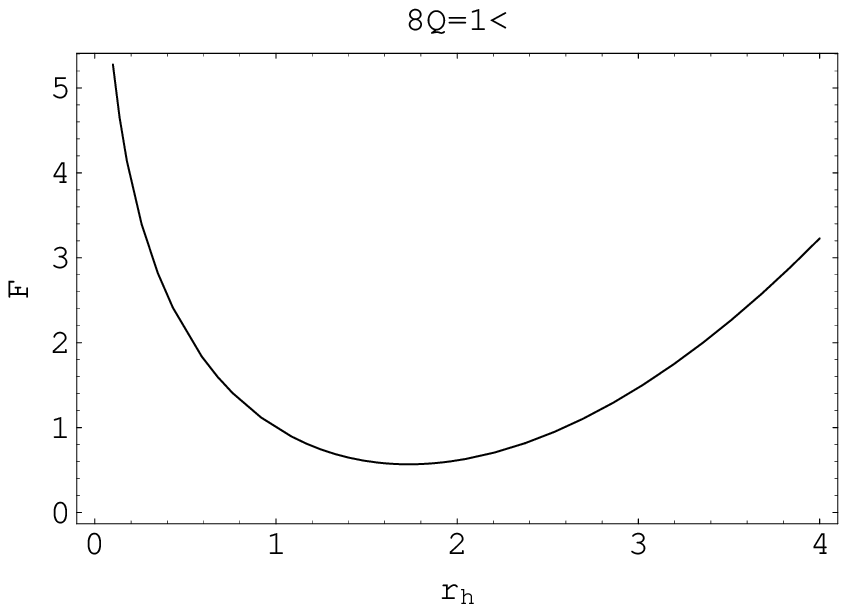}\\
Fig.4(a)~~~~~~~~~~~~~~~~~~~~~~~~~~~~~~~~~~~~~~Fig.4(b)~~~~~~~~~~~~~~~~~~~~~~~~~~~~~~~~~~~~~~~~Fig.4(c)\\
\includegraphics[height=2in, width=2in]{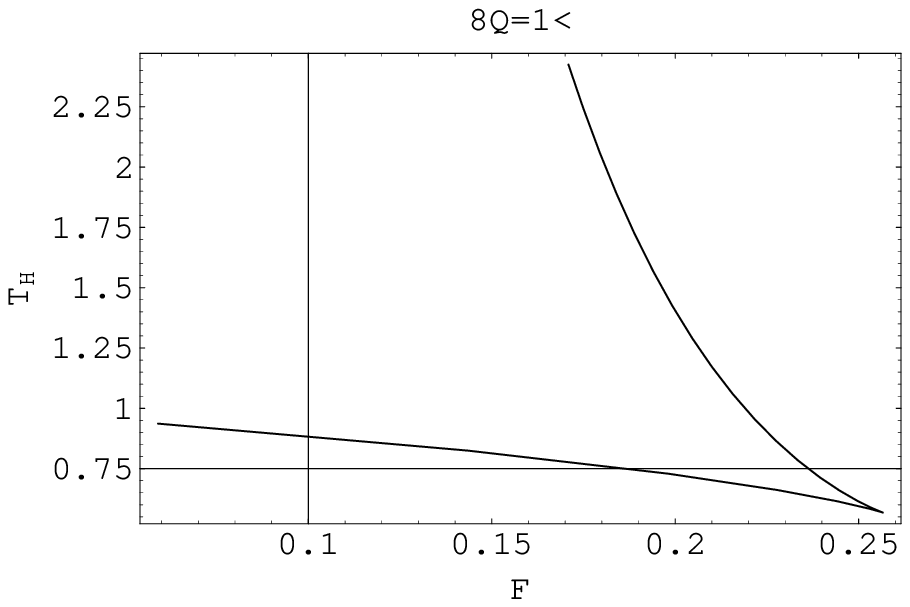}\\

Fig.4(d)\\

\vspace{5mm} Fig. 4(a), 4(b) and 4(c) show the variation of
$T_{H}$, $c_{Q}$ and $F$ respectively with $r_{h}$. 4(d) shows the
variation of $F$ with $T_{H}$ for $\alpha =0.1$ in the case of
EYMGB black hole. \hspace{1cm}
 \vspace{6mm}
\end{figure}

The figures $5(a)-(d)$ show the graphical representation of the
thermodynamical quantities $T_{H}$, $c_{Q}$, $F$ with $r_{h}$ and
also of $F$ with $T_{H}$ for the coupling parameter
$\alpha=0.1~~and~~ Q=1$. From the graphs $T_{H}$ increase sharply
for small $r_{h}$ reaches a maximum value $T_{max}$(at
$r_{h}=\sqrt{3}Q$) and then gradually decreases for large $r_{h}$.
The specific heat is initially positive (for $r_{h}>Q$), gradually
increases and then blows up at $r_{h}=\sqrt{3}Q$ and subsequently
changes sign and remains negative. The free energy is positive
throughout with a minimum at $r_{h}=\sqrt{3}Q$. The variation of
$F$ w.r.t. $T_{H}$ shows a double point at $T_{H}=T_{max}$ which
is of cusp type. this corresponds to a phase transition in BH
thermodynamics. As specific heat changes from positive value to
negative value due due to this phase transition so a stable BH for
small $r_{h}$ becomes a unstable one due to this phase transition.

\section{conclusion}
The thermodynamics of the BHs presented above show some possible
phase transition where the BH changes changes from some stable
configuration to a unstable one and vice versa. At these critical
points the heat capacity changes sign and correspond to
singularities of $c_{Q}$. In EMGB BH for $\lambda=-1$ and $Q=0.5$
there are two critical points at which $c_{Q}$ changes sign and
blows up, indicating a possible phase transition. Further, for the
second critical point there is a transition of $(c_{Q},~F)$ from
$(-,~+)$ sign to $(+,~-)$ sign, i.e., intermediate unstable BH
becomes a stable one and this phase transition is of Hawking-Page
type. However, for the EYMGB BH as $F$ is positive through so
possible no second order phase transition (of Hawking-Page type)
is possible. For further work, it will be interesting to examine
whether Bekenstein area-entropy relation that we have used, should
be modified or not.\\\\

{\bf Acknowledgement:}\\

A part of the work is done during a visit of RB to IUCAA. RB is thankful to IUCAA for warm hospitality and
facilities of research and to State Govt. of West Bengal, India for awarding JRF.\\\\

{\bf References :}\\

$[1]$ Hawking, S. W. : {\it Commun. Math. Phys.} {\bf43}, 199(1975).\\

$[2]$ Bekenstein, J. D. : {\it Phys. Rev. D} {\bf 7}, 2333(1973).\\

$[3]$ Bardeen,J. M. , Carter, B. , Hawking,  S. W. : {\it Commun.
Math. Phys.} {\bf31}, 161(1973).\\

$[4]$ Hut, P. : {\it Mon. Not. R. Astron Soc.} {\bf180}, 379(1977).\\

$[5]$ Davies, P. C. W. : {\it Proc. Roy. Soc. Lond.~A} {\bf 353},499(1977).\\

$[6]$ Davies, P. C. W. : {\it Rep. Prog. Phys.} {\bf41},1313 (1977).\\

$[7]$ Davies, P. C. W. :  {\it Class. Quant. Grav.} {\bf 6}, 1909(1989).\\

$[8]$ Gros, D.J., Perry, M. J., Yaffe, L. G. :  {\it Phys. Rev. D.} {\bf 25}, 330(1982).\\

$[9]$ York, J. W. :  {\it Phys. Rev. D.} {\bf 33}, 2092(1986).\\

$[10]$ Myung, Y. S. :  {\it Phys. Rev. D.} {\bf 77}, 104007(2008).\\

$[11]$ Myung, Y. S. :  {\it Phys. Lett. B} {\bf 645}, 639(2007).\\

$[12]$ Hawking, S. W., Page, D. N. :  {\it Commun. Math. Phys.} {\bf 87}, 577(1983).\\

$[13]$ Brown, J. D., Creighton, J., Mann, R. B. :  {\it Phys. Rev. D} {\bf 50}, 6394(1994).\\

$[14]$ Witten, E. :  {\it Adv. Theor. Math. Phys.} {\bf 87}, 577(1983).\\

$[15]$ Boulware, D. G , Deser, S. :  {\it Phys. Rev. Lett.} {\bf 55}, 2656(1985).\\

$[16]$ Wilthire, D. L. :  {\it Phys. Letts. B} {\bf 169}, 36(1986).\\

$[17]$ Wilthire, D. L. :  {\it Phys. Rev. D} {\bf 38}, 2445(1988).\\

$[18]$ Thibeault, M. , Simeone, C., Eirod, E. F. :  {\it Gen. Rel. Grav.} {\bf 38}, 1593(2006).\\

$[19]$ Chakraborty, S. , Bandyopadhyay, T. :  {\it Class. Quant. Grav.} {\bf25}, 245015(2008).\\

$[20]$ Biswas, R. , Chakraborty, S. : {\it Gen. Rel. Grav.} {\bf DOI : 10.1007/s10714-009-0907-6}\\

$[21]$ Mazharimousavi, S. H., Halilsoy, M. : {\it Phys. Rev. D} {\bf
76}, 087501(2007).\\

$[22]$ Aman, J., Bengtsson, I., Pidocrajt, N. : {\it General. Relativ. Gravit} {\bf
35}, 1733(2003).\\
\end{document}